\documentclass[aps,prd,notitlepage]{revtex4-1}
\pdfoutput=1 
\usepackage{graphicx,color}
\usepackage{amsfonts,amssymb,amsmath}
\usepackage{color}

\begin{document}

\title{Vacuum stability and the scaling behaviour of the Higgs-curvature coupling }

\author{Ian G.\ Moss}
\affiliation{School of Mathematics and Statistics, Newcastle University, 
Newcastle Upon Tyne, NE1 7RU, U.K.}

\begin{abstract}
Stability of the Higgs vacuum during early universe inflation is dependent
on how the Higgs couples to the spacetime curvature. Limits on the curvature
coupling parameter at the electroweak scale $\xi_{EW}$ are shown to be consistent 
only when quantum gravity effects are included and a covariant quantisation procedure, 
such as Vilkovisky-DeWitt, is adopted. Stability requires $\xi_{EW}>-0.03$ for
a top quark mass $M_t=173.34\,{\rm GeV}$.
\end{abstract}


\maketitle

Extrapolation of the Standard Model of particle physics to high energies leads to
the remarkable conclusion that our vacuum is only a long-lived metastable
state, in which the Higgs field sits at a local minimum of the Higgs potential
surrounded by a potential barrier of width somewhere in the range $10^{10}-10^{14}{\rm GeV}$
\cite{Degrassi:2012ry}. 
This raises an interesting question about initial conditions, because, if the Standard Model is 
correct at these energies, then somehow the Higgs field had to 
evolve into the metastable vacuum state during the early stages of the universe
\cite{Espinosa:2007qp}.

In this energy range, there is no reason to abandon General Relativity as the description
of gravity. The focus of this paper is with the effects of a coupling $\xi R{\cal H}^\dagger{\cal H}$, 
between the Higgs field ${\cal H}$ and the curvature $R$, on the evolution of the Higgs field during a period
of early universe inflation. The inflation is driven by an inflaton field, which is assumed weakly
interacting and makes no contribution to the Higgs potential. The curvature coupling increases the height of the 
potential barrier around the metastable minimum if $\xi R$ is positive, and has the opposite effect 
when $\xi R$ is negative, making Higgs stability sensitive to the value of $\xi$. 

The form of the Higgs potential in the range of interest is strongly influenced by quantum effects,
and these are best dealt with using renormalisation group methods. The renormalisation group
uses running coupling constants, whose evolution with energy depends on a set of $\beta$ functions.
In this paper we will see that previous results on $\beta_\xi$ 
\cite{DeSimone:2008ei,Barvinsky:2009ii}
should have quantum gravity corrections even when the Higgs field is small compared to the Planck mass. 
Furthermore, these corrections depend on how the quantum field theory is constructed. Clearly, a 
unique result is desirable, and for this an approach based on the principle of
covariance under field transformations will be adopted \cite{DeWittdynamical}. 
Covariant approaches are widely used for non-linear sigma 
models \cite{AlvarezGaume198185}, but their importance for Higgs physics has been
relatively unappreciated..

The field transformation used most frequently for Higgs cosmology is a conformal re-scaling
of the metric which removes the curvature-coupling term, transforming the theory from the
original Jordan frame to the Einstein frame. It has been pointed out before that quantum 
calculations can lead to different results when done in the Einstein frame instead of the 
Jordan frame, and covariant approaches have already been proposed to resolve 
inconsistencies \cite{Kamenshchik:2014waa,George:2015nza}.
We shall see that covariant quantisation gives consistent results on Higgs instability, 
and the results differ from those obtained previously \cite{Herranen:2014cua}. 
The last thing we want to see is a Higgs field 
which is unstable in the Jordan frame and stable in the Einstein frame.

The scaling behaviour of Higgs couplings can be inferred from close look at the
effective potential. The Higgs effective potential is written as a function 
$V_{\rm eff}(\xi,\lambda,\phi,\mu)$, where $\xi$ and $\lambda$ are running
couplings depending on $\mu$, the renormalisation scale. At one loop order,
the explicit dependence on renormalisation scale has contributions from
each of the fields which couples to the Higgs. These contributions are determined by
a set of second order operators $\Delta^n(\phi)$. The $\beta$ functions are
obtained by comparing coefficients in the renormalisation group equation
\cite{Sher:1988mj},
\begin{equation}
\beta_\xi{\partial V_{\rm eff}\over\partial\xi}+
\beta_\lambda{\partial V_{\rm eff}\over\partial\lambda}
-\gamma\phi{\partial V_{\rm eff}\over\partial\phi}
={1\over 16\pi^2}\sum_n (\pm) b_2(\Delta_n)\label{scaling},
\end{equation}
where the sign is positive for bosons and negative for fermions and ghosts.
Renormalisation of $\phi$ is responsible for the anomalous dimension $\gamma$.
The coefficients $b_2$ are known for most types of operators on
arbitrary spacetime backgrounds (e.g. \cite{Vassilevich:2003xt}). 
For example, the wave operator with mass
${\cal M}(\phi)$
has $b_2(-\nabla^2+{\cal M}^2)={\cal M}^4/2-R{\cal M}^2/12+O(R^2)$.

With the inclusion of gravity, the feature which we need to
focus on is the fact that the field space develops a non-trivial geometry, as
seen particularly in the non-linear sigma models. Covariant approaches to quantum field theory,
such as the Vilkovisky-DeWitt formalism 
\cite{Vilkovisky:1984,DeWittdynamical,barvinsky1985generalized,ParkerTomsbook}, 
take advantage of this field-space geometry.
In the general case of a gauge theory with fields $\varphi^a$ and action $S[\varphi^a]$,
the field operator is given by
\begin{equation}
\Delta_{ab}={\delta^2 S\over\delta\varphi^a\delta\varphi^b}
-\Gamma^c{}_{ab}{\delta S\over\delta\varphi^c}
+{1\over 2\alpha}K^a{}_\epsilon[\varphi]K_b{}^\epsilon[\varphi],
\label{vdw}
\end{equation}
The innovation of Vilkovisky and DeWitt was to put the second functional derivatives into covariant form
by introducing a connection $\Gamma^a{}_{bc}$. The connection ensures that the effective action is 
covariant under field redefinitions. It can be disregarded when the background field is 
on-shell, i.e. $\delta S/\delta\varphi^a=0$, but it has to be included when calculating the $\beta$ 
functions from the effective action. The final term in (\ref{vdw}) is a gauge-fixing term for a 
gauge-fixing functional $\chi_\epsilon=K^a{}_\epsilon\delta\varphi_a$. In the Landau gauge 
limit $\alpha\to 0$, the connection reduces to the Levi-Civita connection for the metric on the space of fields.
Due to this simplification of the connection, all the results below are calculated using
Landau gauge, although the Vilkovisky-DeWitt result is independent of the choice of gauge
fixing.

The gravity gravity-scalar sector consists of a spacetime metric $g_{\mu\nu}$ and the Higgs
doublet field ${\cal H}$. For convenience, we can replace the Higgs doublet by a set of 
four real scalars $\phi^i$ with potential $V(\phi^i)=\lambda\phi^4/4$.  The Lagrangian density for the gravity-Higgs 
sector ${\cal L}_g$ is,
\begin{equation}
{\cal L}_g={1\over 2}M_p^2U(\phi)R\,|g|^{1/2}
-\frac12 G_{ij}(\phi)\,g^{\mu\nu}\partial_\mu\phi^i\partial_\nu\phi^j\,|g|^{1/2}
-V(\phi)\,|g|^{1/2},\label{nlsm}
\end{equation}
where $\partial_\mu$ denotes an ordinary spatial derivative and $8\pi G M_p^2=1$. The non-minimal 
coupling terms are contained in the function $U(\phi)$ multiplying the Ricci scalar $R$, 
$U(\phi)=1-\xi\phi^2/M_p^2$.
Superficially, the $\xi$ term resembles a contribution $\xi R$ to the Higgs mass, but this is not the
whole story when we consider the one-loop corrections.

Variations in field space can be combined into fields $\eta^a=(\delta g_{\mu\nu}/2\kappa,\delta\phi^i)$, scaled
so that $\eta^a$ has the dimensions of mass. The total operator takes the form,
\begin{equation}
\Delta{}_{ab}=-{\delta^2 S\over\delta\eta^a\delta\eta^b}+
\Gamma^c{}_{ab}{\delta S\over\delta\eta^c}=
-{\cal G}_{ab}\nabla^2+\zeta{\cal P}^{\alpha\beta}{}_{ab}\nabla_\alpha\nabla_\beta
+{{\cal M}^2}_{ab},
\end{equation}
where  $\zeta=1-1/\alpha$ is a gauge parameter. There are three important tensors in this expression: ${\cal G}_{ab}$ 
is the metric on field space, ${\cal P}^{\alpha\beta}_{ab}$ combines the non-minimal derivative terms, 
and ${{\cal M}^2}_{ab}$ is an effective mass term. (Explicit expressions for ${\cal G}_{ab}$ and
${\cal M}^2{}_{ab}$ can be found, for example in Ref.
\cite{Barvinsky:2009fy}, but in the ubiquitous Feynman gauge $\zeta=0$). The metric is
used to construct the 
Levy-Civita connection in (\ref{vdw}) 
by the usual expression,
\begin{equation}
\Gamma^a{}_{bc}=\frac12 {\cal G}^{ad}\left({\delta {\cal G}_{db}\over \delta \eta^c}+
{\delta {\cal G}_{dc}\over \delta \eta^b}-{\delta {\cal G}_{bc}\over \delta \eta^d}\right).
\end{equation}

The operator simplifies considerably for constant background values of the Higgs field which
are also below the Planck scale, $|\xi|^{1/2}\phi\ll M_p$. In this case the gravity-scalar cross
terms drop out of the operator, and the Higgs effective mass term reduces to
\begin{equation}
{{\cal M}^2}_{ij}=\frac12 G_{ij}\,R+V_{;ij}.\label{mass}
\end{equation}
Note, in particular, that the $\xi R$ mass term has been cancelled by the Vilkovisky-DeWitt
corrections. The contribution to $\beta_\xi$ can be read off from the quadratic terms in (\ref{scaling}). 
The anomalous dimension $\gamma$  does not contribute because it has no terms of order $\lambda$,
and we get
\begin{equation}
\beta_\xi=
\begin{cases}
4\lambda&\mbox{covariant}\\
2\lambda(6\xi-1)&\mbox{non-covariant}.
\end{cases}
\end{equation}
The result for $\beta_\xi$ differs substantially from the non-covariant result \cite{Herranen:2014cua}, 
which is has the $\xi R$ mass term. 
How can the inclusion of quantum gravity make such a difference? The underlying reason is that 
requiring covariance under 
field redefinitions means that the metric and Higgs fields can no longer be treated separately, 
and quantising one but not the other is inconsistent. From a technical point of view, the
quantum gravity effects on $\xi$ survive in the $M_p\to \infty$ limit due to the 
$\Gamma^{\mu\nu}{}_{ij}$ connection components \footnote{In non-linear sigma 
model terms, the covariant
metric expansion is 
$\delta g_{\mu\nu}/(2\kappa)=\eta^{\mu\nu}+\Gamma^{\mu\nu}{}_{ij}\eta^i\eta^j$.}.

In order to verify the covariance, consider the same calculation in the Einstein frame, 
where the metric $\hat g_{\mu\nu}=Ug_{\mu\nu}$ and the curvature term reduces to
$\hat U=1$. The scalar field-space metric becomes
\begin{equation}
\hat G_{ij}=U^{-1}G_{ij}+6\xi^2(M_pU)^{-2}\phi_i\phi_j.
\end{equation}
This time there are no $\xi R$ mass terms anywhere. The Vilkovisky-DeWitt corrections \cite{Moss:2014nya} 
provide the connection terms for $V_{;ij}$ in (\ref{mass}), but do little else and we recover exactly the same 
result for $\beta_\xi$ as before,
\begin{equation}
\beta_\xi=
\begin{cases}
4\lambda&\mbox{covariant}\\
-2\lambda&\mbox{non-covariant}.
\end{cases}
\end{equation}
In fact, we can see directly that $\xi$ always appears in the combination 
$\xi\kappa$ and has to drop out of the $\beta$ functions in the $M_p\to\infty$ limit.

The fermion and gauge boson contributions to the $\beta$ functions are simpler because
they have no background values and therefore there is no mixing with metric fluctuations
in the field operators. The Vilkovisky-DeWitt corrections are important nevertheless, 
and give the effective mass terms in table \ref{tab}. Note that the vector 
boson and ghost masses are 
equal when the Vilkovisky-DeWitt corrections are included. This solves a problem in the
Standard Model on a curved background without corrections, where the ghosts do 
not precisely cancel the unphysical gauge modes.

\begin{table}[ht]
\caption{Corrected effective mass terms for the standard model particles in curved spacetime.}
\vskip 1mm
\centering
\begingroup
\renewcommand{\arraystretch}{1.2}
\begin{tabular}{lccl}
\hline
Field&Components&Sign&Square mass ${\cal M}^2$\\ [0.5ex]
\hline
$t$&$12$&$-$&$\frac12y^2\phi^2+\frac14 R$\\
$W^\pm$&8&$+$&$\frac14g^2\phi^2+\frac12R$\\
$W^\pm$(ghost)&2&$-$&$\frac14g^2\phi^2+\frac12R$\\
$Z^0$&4&$+$&$\frac14(g^2+g^{\prime 2})\phi^2+\frac12R$\\
$Z^0$(ghost)&1&$-$&$\frac14(g^2+g^{\prime 2})\phi^2+\frac12R$\\
\hline
\end{tabular}
\label{tab}
\endgroup
\end{table}

In the $|\xi|^{1/2}\phi \ll M_p$ limit, the anomalous dimension is
given by
\begin{equation}
16\pi^2\,\gamma=3y^2-\frac94g^2-\frac34g^{\prime\,2}.
\end{equation}
Substituting the masses into (\ref{scaling}) and comparing the quadratic terms gives
\begin{equation}
16\pi^2\,\beta_\xi=4\lambda+
\xi\left(6y^2-\frac32g^{\prime\,2}-\frac92 g^2\right)
-y^2+\frac12g^{\prime \,2}+\frac32g^2.\label{betaxi}
\end{equation}
All other $\beta$ functions are the same as in flat space.

The result for $\beta_\xi$ contradicts previous work, e.g. Ref. 
\cite{DeSimone:2008ei,Herranen:2014cua} which give the non-covariant result.
There are a number of contradictory results in the large field limit $|\xi|^{1/2}\phi\sim M_p$.
Ref. \cite{Bezrukov:2009db} gives results for $\beta_\xi$ without quantum gravity corrections,
whilst Ref. \cite{Barvinsky:2009fy} includes gravity but omits Vilkovisky-DeWitt corrections.
The closest comparison is to results in the Einstein frame using
covariant methods on the scalar (but not the gravity) sector \cite{George:2015nza}, 
$\beta_\xi=\gamma\xi$, which agrees with (\ref{betaxi}) at large $\xi$.

\begin{figure}[htb]
\begin{center}
\includegraphics[width=0.4\textwidth]{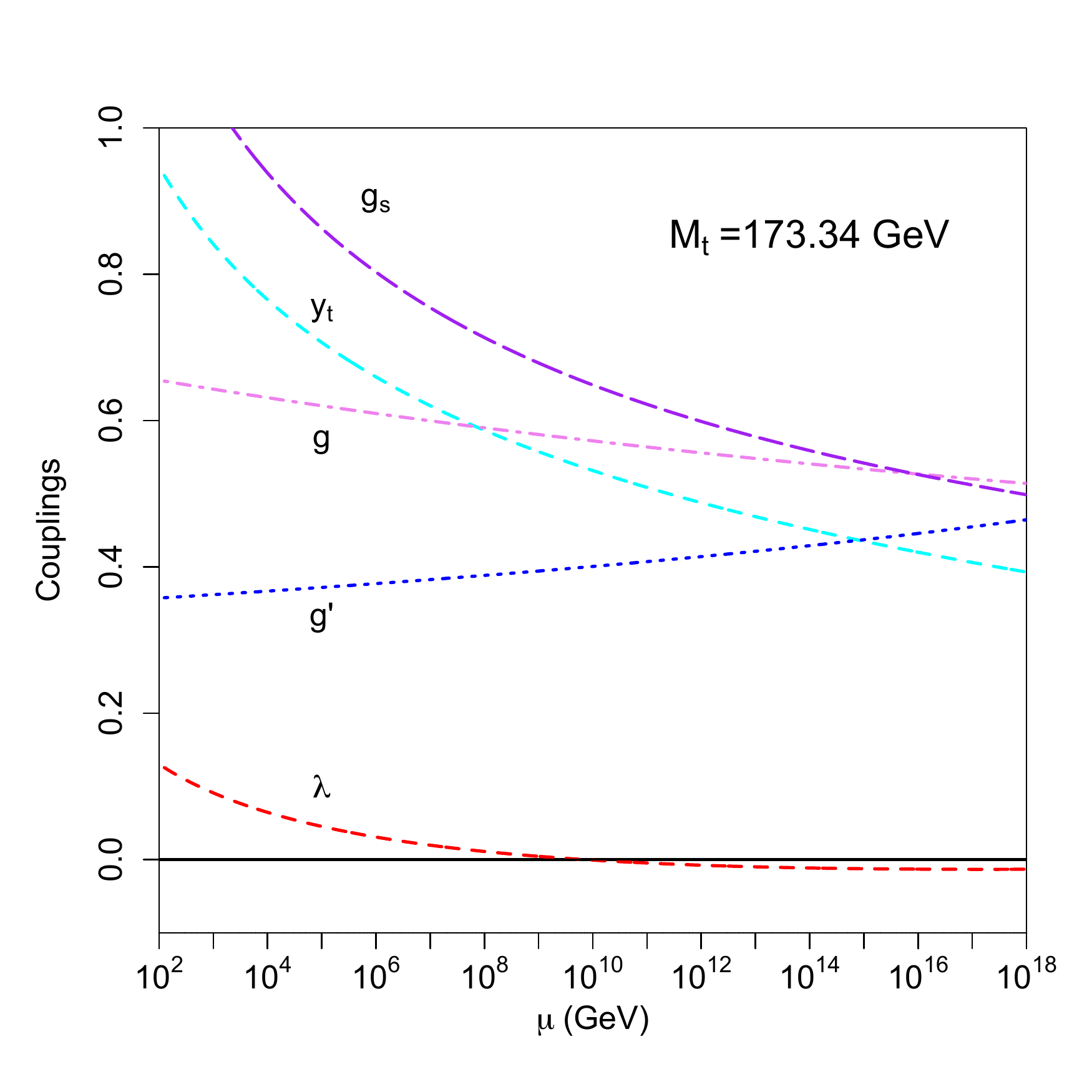}
\includegraphics[width=0.4\textwidth]{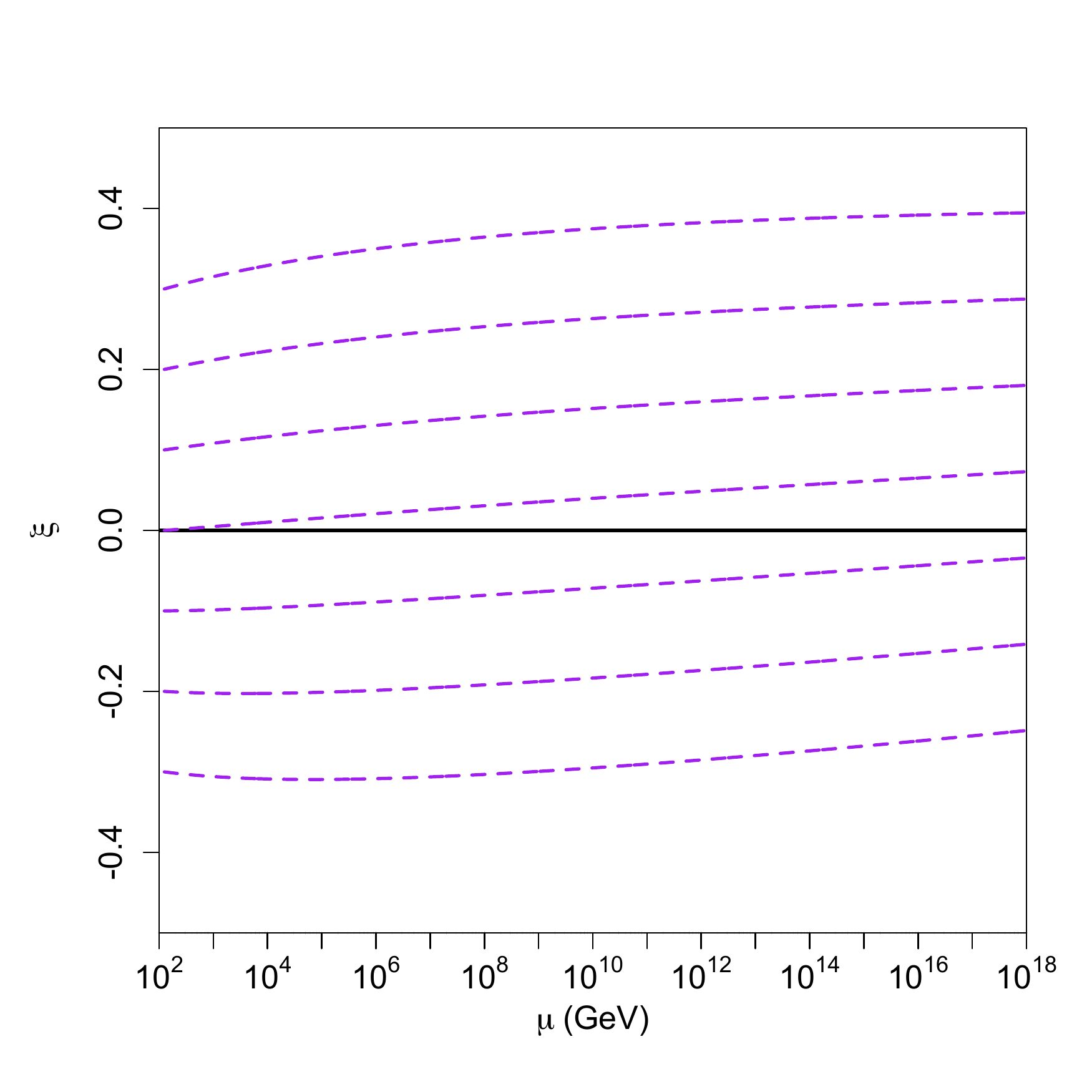}
\caption{
Running couplings with $m_{\rm top}=173.34{\rm GeV}$. Two-loop $\beta$
functions have been used for the standard model couplings.}
\label{couplings}
\end{center}
\end{figure}

The $\beta$ functions are effectively a set of differential equations which can be solved
to obtain the dependence of the couplings on the renormalisation scale. This has been 
done in figure \ref{couplings}, using the flat-space two-loop $\beta$ functions for the 
couplings on the left hand plot \cite{Degrassi:2012ry}. Since the one-loop results work well 
enough for all of the 
couplings apart from $\lambda$, it is reasonable to use the one-loop result for $\beta_\xi$ 
which has been derived above. The running coupling $\xi(\mu)$, shown in the right-hand plot, 
displays very little change with energy, apart from a small increase of around $0.07$ from the 
Electroweak to the Planck scale. 

\begin{figure}[htb]
\begin{center}
\includegraphics[width=0.4\textwidth]{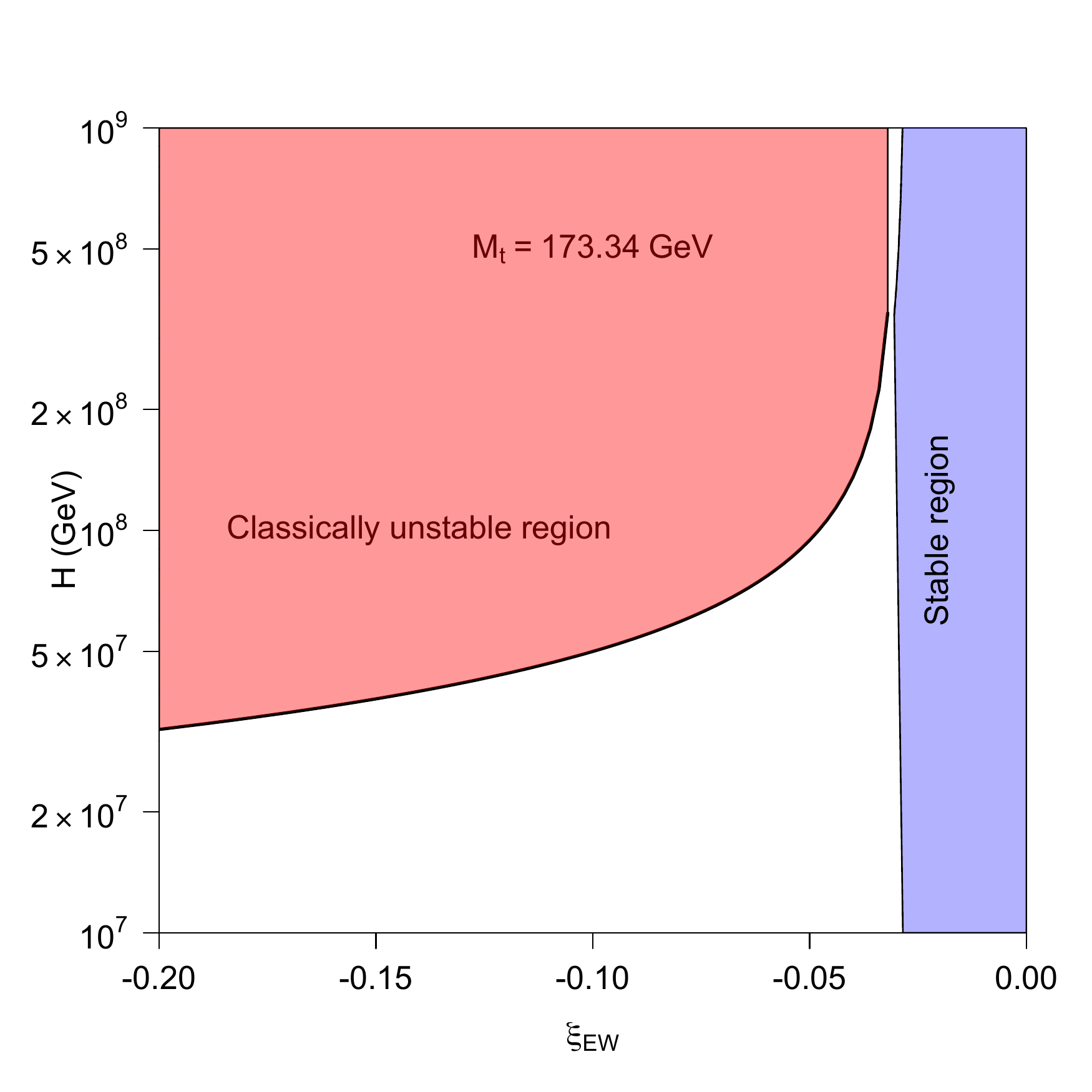}
\caption{
The inflationary Higgs instability region and the lower limit on $\xi_{EW}$ as a function
of the Hubble parameter during inflation.}
\label{range}
\end{center}
\end{figure}

The running couplings can be combined with the field renormalisation
to construct an improved form of the effective potential,
\begin{equation}
V_{\rm eff}=\frac12\xi_{\rm eff}(\phi)R\,\phi^2+\frac14\lambda_{\rm eff}(\phi)\,\phi^4,
\end{equation}
where the effective couplings combine the running coupling with the field renormalisation
at $\mu=\phi$ \footnote{There is a proviso 
here that the effective mass terms ${\cal M}^2$ must all be larger than the curvature.}.
If $\lambda_{\rm eff}$ is negative at large $\phi$, then negative values for $\xi$ tend to 
destabilise the Higgs field during an early universe
inflationary period. For some critical value $R_c$, the potential develops an inflection
point and the field become classically unstable for $R>R_c$. The inflection 
point occurs when $V'=V''=0$, which can be solved using the running couplings to 
obtain a relation between $R_c$ and the values of the couplings at the electroweak scale.
Since most of the couplings are known, in effect we have a relation between $R_c$ and
the value of the curvature coupling measured at the electroweak scale, $\xi_{EW}$.
The region of instability for negative $\xi_{EW}$ is shown in figure \ref{range},
in terms of the Hubble parameter $H$, where $R=12 H^2$. Curvature couplings $\xi_{EW}<-0.32$ 
can occur, but only in models of inflation which have unusually
small values of the Hubble parameter.

Values of $\xi_{EW}>-0.32$ have the opposite effect and stabilise the Higgs field
against decay associated with quantum tunnelling through the potential barrier.
The probability of fluctuations through the potential barrier is exponentially suppressed
with exponent $8\pi^2 V(\phi_c)/3H^4$ \cite{Espinosa:2007qp}. Figure \ref{range}
shows the parameter region for $8\pi^2 V(\phi_c)/3H^4>10$ and leads to the stability
bound $\xi_{EW}>-0.3$. (Using a lower top quark mass $M_t=173\,{\rm GeV}$ raises
the values of the Hubble parameter but leaves the limit on $\xi_{EW}$ unchanged.)

In summary, stability of the Higgs vacuum during early universe inflation sets limits on the
curvature coupling. These limits have been shown to be frame-independent in the
$|\xi|^{1/2}\phi\ll M_p$ limit, but only when  quantum gravity effects are included and a 
covariant quantisation procedure such as Vilkovisky-DeWitt, is adopted. We finish with some
brief remarks on on extending the renormalisation group calculations
to the regime $|\xi|^{1/2}\phi\sim M_p$. This is technically feasible with the covariant
approach, but moving to larger values of the
Higgs field makes it necessary to consider the effects of higher order operators on the
renormalisation group, and because of this predictability begins to be eroded
\cite{Burgess:2014lza}. On the positive side, these higher border terms can stabilise the 
Higgs vacuum and remove the need for a separate field driving inflation.

\section*{Acknowledgements} 
The author would like to thank David Toms and Gerasimos Rigopoulos for helpful discussions
during the preparation of this paper, and the STFC for financial support 
(Consolidated Grant ST/J000426/1).

\bibliography{paper.bib}
\end{document}